\documentclass[preprint,5p,authoryear]{elsarticle}

\usepackage{lineno,hyperref}
\modulolinenumbers[5]
\usepackage{amssymb}
\usepackage{comment}

\journal{Cryogenics}
\begin{document}


\begin{frontmatter}

\title{The Simons Observatory: Improved Cryogenic Struts for use in the Large Aperture Telescope Receiver}

\author[UPenn]{John Orlowski-Scherer\corref{corauth}}
\ead{jorlo@sas.upenn.edu}
\cortext[corauth]{corresponding author}
\author[UChicago,Kavli]{Anna Kofman}
\author[UPenn]{Tanay Bhandarkar}
\author[UPenn]{Mark Devlin}
\author[UPenn]{Saianeesh K. Haridas}
\author[UPenn]{Jeff Iuliano}
\author[UPenn]{Alex Manduca}
\author[UPenn]{Robert J. Thornton}

\address[UPenn]{Department of Physics and Astronomy, University of Pennsylvania, 209 South 33rd Street, Philadelphia, PA, 19104, USA}
\address[UChicago]{University of Chicago, Department of Astronomy and Astrophysics, 5720 S Ellis Ave, Chicago, IL, 60637, USA}
\address[Kavli]{Kavli Institute for Cosmological Physics, University of Chicago, 5640 S Ellis Ave, Chicago, IL, 60637, USA}

\begin{abstract}
The Simons Observatory Large Aperture Telescope Receiver (SO LATR) is a next generation Cosmic Microwave Background camera equipped with $>60,000$ detectors operating at $100$\,mK. Maintaining these detectors at the correct temperatures and locations requires stiff, cryogenically insulating struts. In this paper we report the design and performance of a novel glue joint in a strut used in the SO LATR to achieve the required performance. We use a tapped hole and set screw to create a profile on the exterior and interior wall of the glue joint, respectively, which greatly increases the strength of that joint by changing the failure mode from adhesive to cohesive. The failure mode of the resulting glue joint is cohesive with a yield strength $10\%$ higher than a comparable smooth-walled design, and an ultimate strength $33\%$ higher. Comparisons of the measured yield strength to the predicted axial load on the strut from simulations results in a factor of safety for the strut of 7. These struts have been installed in the SO LATR for three years and have undergone numerous thermal cycles from $300$\,K to $100$\,mK with no evidence of damage to the glue joint.
\end{abstract}


\end{frontmatter}

\section{Introduction}
The Simons Observatory \citep[SO;][]{SOgoals} is a next-generation Cosmic Microwave Background (CMB) experiment, which recently began observations from its $5200$\,m elevation site on Cerro Toco in northern Chile.  The observatory consists of three Small Aperture Telescopes \citep[SATs;][]{Galitzki2024} and one Large Aperture Telescope telescope \citep[LAT;][]{Xu2021}.  The LAT will make an arcminute resolution map of approximately 60\% of the sky in both temperature and polarization, which will advance many priorities in cosmology -- for example by precisely measuring the damping tail of the CMB power spectrum and identifying galaxy clusters as tracers of mass in the universe. With over 60,000 Transition Edge Sensor detectors in the LAT due to the recent Advanced SO upgrade\citep[ASO;][]{ASOgoals} and another 35,000 in the three SATs, the SO maps will rapidly become the deepest of their kind ever made.  

In order to detect the $\mu$K scale fluctuations in the CMB, SO uses transition edge sensors \citep[TESes;][]{mccarrick2021}. TESes work by holding a strip of metal on its superconducting transition. 
The TESes need to operate at cryogenic temperatures, at least $\lesssim 250$\,mK but ideally $\sim100$\,mK. Therefore, one of the primary challenges of the LAT design process was to plan a cryostat large enough to house $>60,000$ ASO TESes while maintaining a stable base temperature of $\leq100$\,mK. The result of that design process is the LAT Receiver \citep[LATR;][]{Zhu2021, orlowski2018}.

The receiver is split into multiple temperature stages -- $300$\,K, $80$\,K, $40$\,K, $4$\,K, $1$\,K, and $100$\,mK -- which are cooled via Pulse Tube (PT; $80, 40, 4$\,K) coolers and a Dilution Refrigerator (DR, $1$\,K and $100$\,mK). Within the LATR, the detectors are housed in $13$ separate optics tubes (OTs), which contain reimaging optics and filters in addition to the detectors. The OTs contain three temperature stages, $4$\,K, 1\,K, and $100$\,mK. The DR is thermally connected to the 1\,K and 100\,mK stages in the OTs via the thermal back-up-structure (BUS). The thermal BUS consists of two copper plates, one at $1$\,K and one at $100$\,mK, which are connected by carbon fiber trusses.


\begin{figure*}
\centering
\includegraphics[width=0.85\paperwidth]{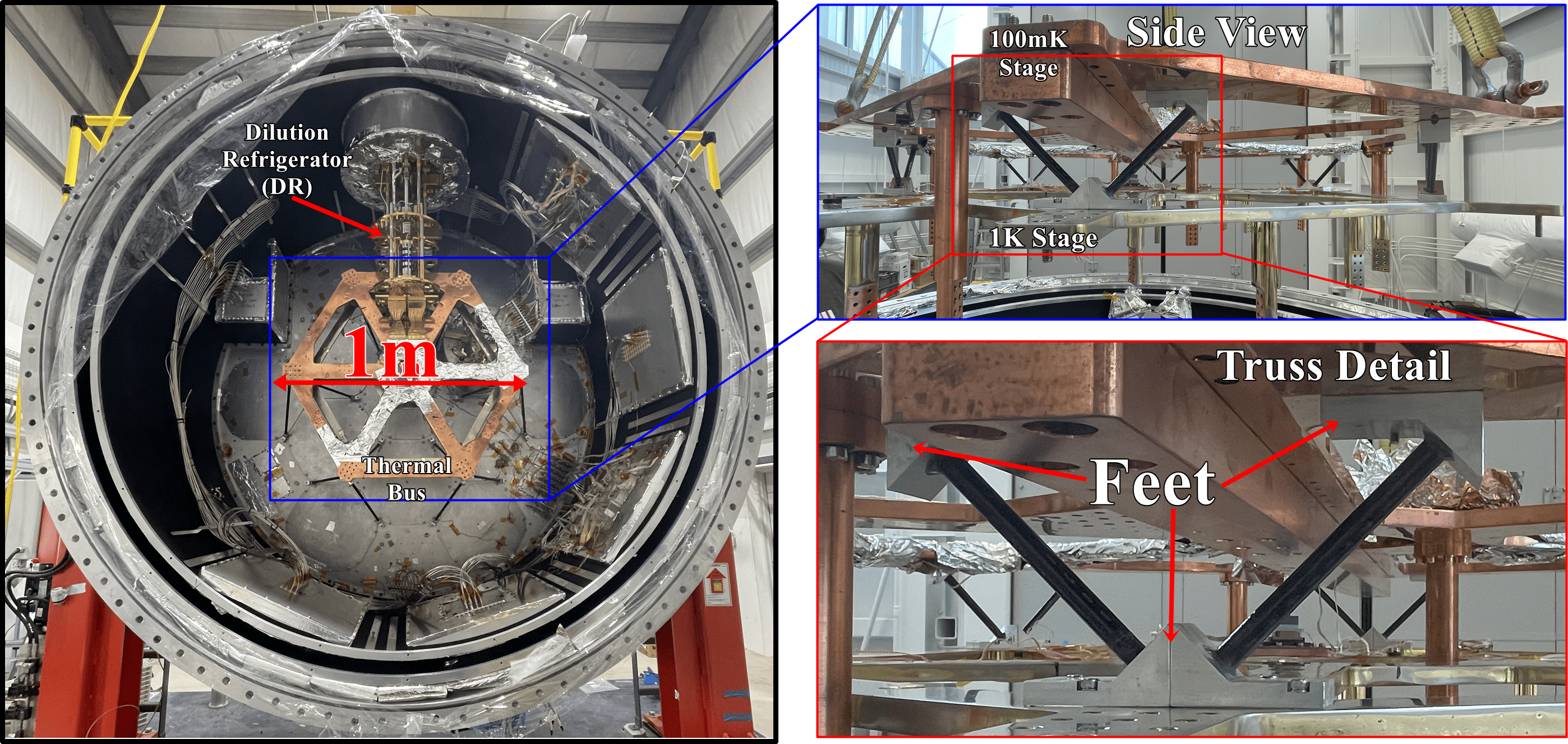}
\caption{Diagram showing the thermal BUS truss in-situ in the LATR. The left image shows the thermal BUS as installed in the LATR. The top right image is a side view of the thermal BUS, showing the distribution of the six trusses around the perimeter of the thermal BUS. The bottom left image is a detail of a strut, showing the feet, carbon fiber rod, and external glue joint. }
\label{fig:truss}
\end{figure*}


This paper details the novel design of the carbon fiber struts used in the construction of the LATR thermal BUS. These struts are constructed by gluing carbon fiber rods into aluminum feet.  In this configuration, the glue joint is generally the first point of failure when loading the truss to its limit\footnote{Throughout, a strut is a single glued carbon fiber and aluminum element, while the truss is the entire assemblage of struts }.  We describe here a novel approach to this glue joint, and related measurements associated with the development of the LATR trusses. Related work was done to develop trusses for the SATs, which is detailed in \citet{Crowley2022}.


This paper is organized as follows. In Section~\ref{sec:design} we describe the design of the struts and in particular of their glue joints. In Section~\ref{sec:sims} we describe the finite element analysis done to predict the structural performance of these struts. In Section~\ref{sec:testing} we describe the testing methodology used to evaluate the performance of these struts, while in Section~\ref{sec:results} we present the results of that testing. Finally, in Section~\ref{sec:conc} we summarize the results of this paper and present some recommendations for future improvements.

\section{Design}
\label{sec:design}
The design of the strut is comprised of the carbon fiber tube itself, glued at both ends into 6061 aluminum ``feet'' which provide a mating interface to the thermal BUS. Figure~\ref{fig:truss} shows a fully assembled truss as installed in the thermal BUS. The novelty in our design lies in the gluing interface between the carbon fiber tube and the feet. For many truss designs, including our own, the limiting factor, in terms of both yield and ultimate strength is not the carbon fiber but the glue joint. The advantage of a truss design is that, to good approximation, each individual member is under only tension or compression, with negligible shear forces. This makes good use of the high tensile strength of carbon fiber \citep[e.g.;][]{Wu2018}; however, it places significant strain on the glue joint, as the forces on the strut are shearing the glue. This may cause the glue to fail, either adhesively, via detachment of the glue from one of the glued surfaces, or cohesively, when the glue itself fractures. 

Our novel design increases the glue joints' strength by adding a profile to the surfaces of the strut feet which are glued to the carbon fiber rods. In early designs for this strut, the gluing interface was simply a cut annulus, slightly larger in outer diameter and smaller in inner diameter than the carbon fiber tube. 
In testing, these simple designs would fail adhesively, short of the cohesive failure limit.  We therefore set out to design a new interface which would fail cohesively, gaining strength in our strut design.  

\subsection{Glue Joint}

\begin{figure}
\centering
\includegraphics[width=0.9\columnwidth]{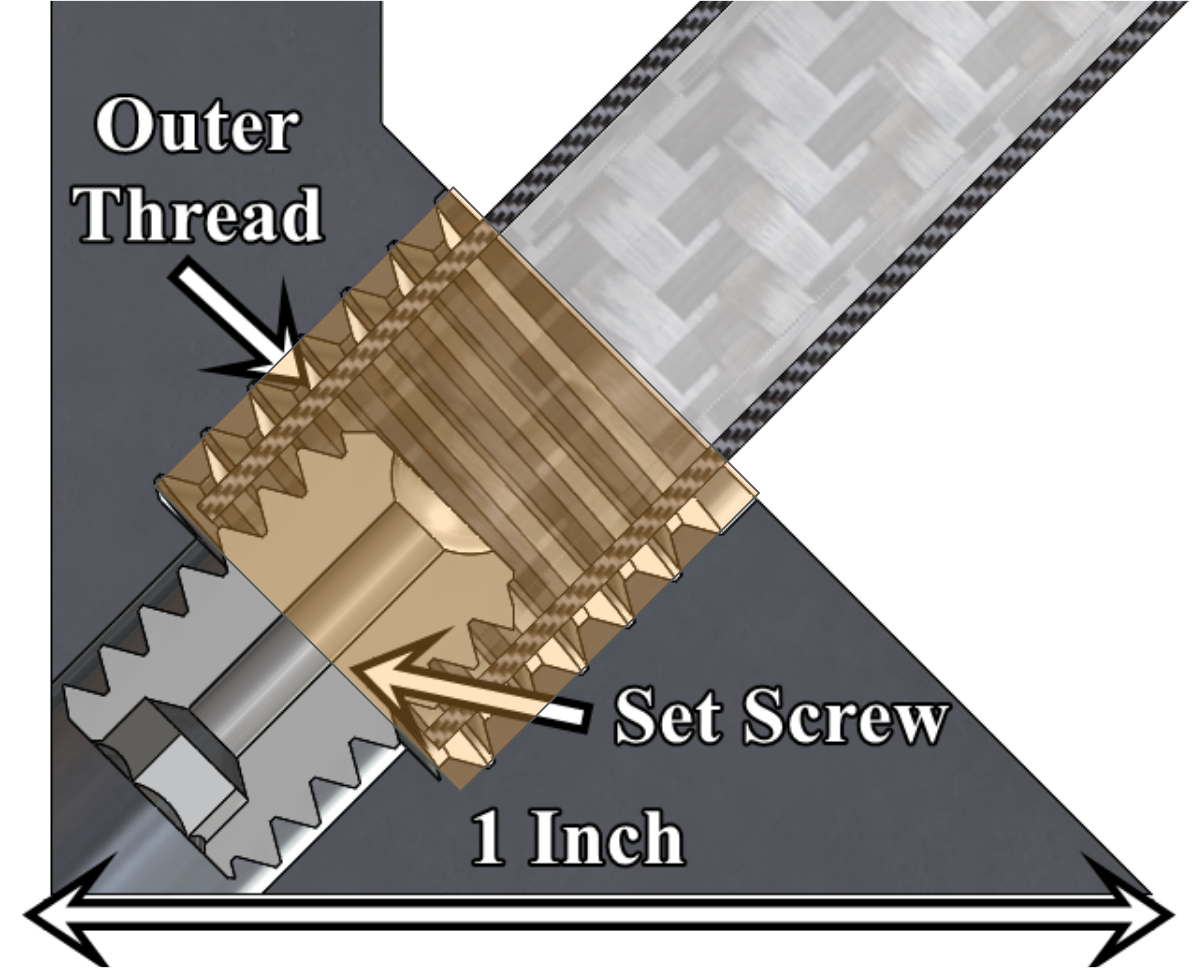}
\caption{A cut away showing the glue interface in our strut design with the tapped outer wall and setscrew inner wall. In our design the outer tap is a M$10\times1.5$ while the inner set screw is a $3/8$ inch long $1/4"-20$ vented cup point. The area the glue fills is indicated with an orange color. The threaded profiles of the inner and outer wall prevent the glue from adhesively failing.}
\label{fig:glue_interface}
\end{figure}

The design we settled on is shown in Figure~\ref{fig:glue_interface}. It is still an annulus, but now the profile of both walls is zig-zagged. This has a double effect: firstly, it reduces the shear force on the glue by a factor of $\cos(\theta)$ where $\theta$ is the angle of the zig-zag. 
Secondly, it has the effect of mechanically constraining the glue so that the glue between two threads must cohesively fail for the joint to fail. To simplify the manufacture of the part, the outer wall profile is cut with a M10x1.5 thread tap, while the inner wall consists of a set screw, which mates to a tapped hole in the base of the foot. We use vented set screws as they eliminate the virtual leak from the carbon fiber tubes under vacuum. We considered other more complicated profiles, such as a pocket, but this design minimized the time and cost of manufacture while failing cohesively as opposed to adhesively. The carbon fiber rods selected were $0.250$ inch ID $0.320$ inch OD twill finish, standard modulus circular rods from Clearwater Composites.\footnote{Clearwater Composites, LLC 4429 Venture Ave. Duluth, MN 55811, https://www.clearwatercomposites.com/}

\subsection{Gluing Process}

The gluing process for the struts that were tested was identical to that used in the production of the actual struts. It is as follows. First, the carbon fiber rods were roughened with sandpaper until water would bead on them. The groove of the smooth-walled feet was also roughened with sandpaper. Both the smooth and profile walled feet were cleaned in an ultrasonic bath, while the carbon fiber rods were cleaned with isopropyl alcohol and were allowed to dry. We then mixed a batch of 3M Scotch-Weld Epoxy Adhesive 2216 in the 5 to 2 weight ratio prescribed by the manufacturer for cryogenic use \citep{3m2018, amils2016, rondeaux2002}. After mixing, the glue was placed in a rough vacuum and pumped on until bubbles stopped coming to the surface, about $10$ minutes. The process then diverges slightly for the smooth and profile-walled struts. 

For the smooth-walled struts, the groove in the foot was filled with glue and a coating of glue was applied to the outer surface of the carbon fiber rod. The rod was then inserted into the foot and twisted to evenly spread the glue. The rod was then removed and the above procedure repeated to ensure the groove was completely filled with glue. The foot was then bolted to the gluing jig, and the procedure was repeated on the other foot, completing the gluing process. 

For the profile-walled struts, first the groove of the foot was filled with glue, with the glue worked into the threading on the outer wall by means of a tongue depressor. Next, glue was applied to the threads of the set screw, again being worked into the threading using a tongue depressor. The set screw was then threaded into the matching threading on the foot until it was flush with the surface of the foot. More glue was injected into the groove in the foot between the outer wall and the set screw until it was full. Glue was then applied to the outer surface of the carbon fiber rod, and it was inserted into the groove and rotated to evenly spread the glue. It was then removed and the above step repeated. The foot of the strut was then bolted to the gluing jig, and the entire process repeated on the other side, completing the gluing process. 

\begin{figure}
\centering
\includegraphics[width=0.9\columnwidth]{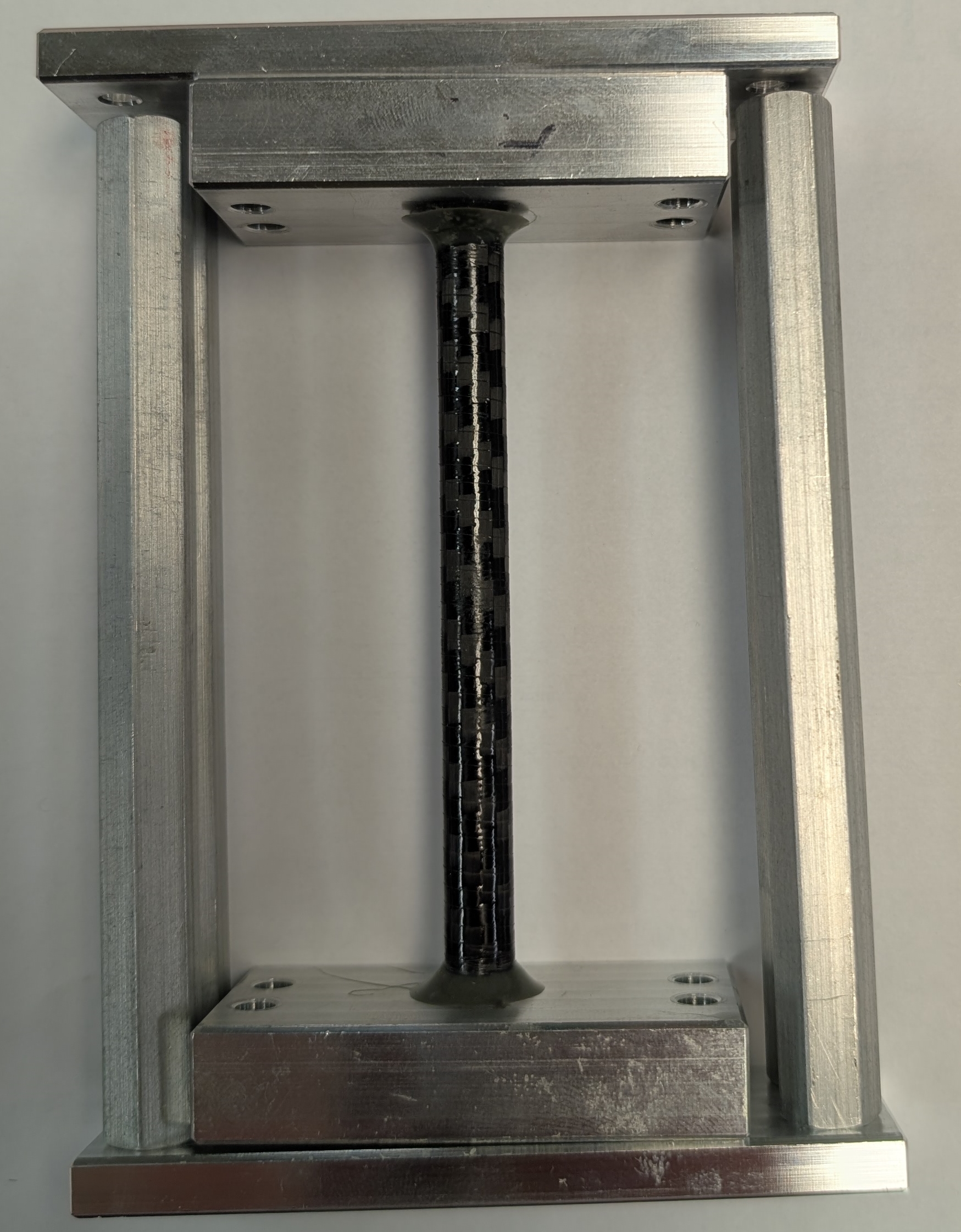}
\caption{One of the sample pieces in the gluing jig. The samples and gluing jig are identical between profile- and smooth-walled except for the annulus shape. }
\label{fig:jig}
\end{figure}

Two struts would be glued at once, and this entire process took less than $45$ minutes, well less than the $90$ minute glue setting time. The glue was allowed to harden overnight, about $16$ hours, with the jigs being rotated every $10$ minutes for the first $2$ hours to ensure the glue did not leak out of the groove under the influence of gravity. After partially curing at room temperature, the parts were baked for $2$ hours at $150-160$F, to finish the glue curing process. See Figure~\ref{fig:jig} to see the jig. 


\section{Simulations}
\label{sec:sims}
\begin{figure}
\centering
\includegraphics[width=0.9\columnwidth]{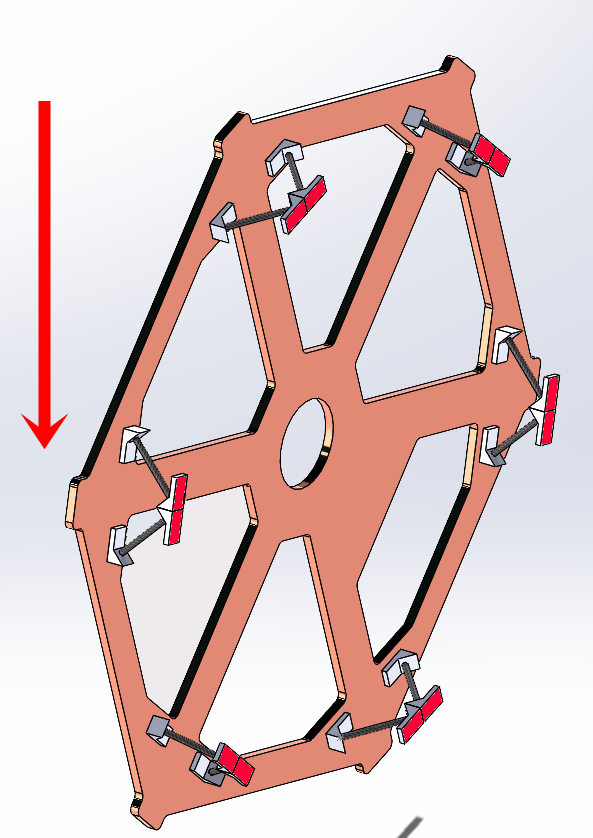}
\caption{Simulation set up with the arrow indicating the direction of the applied load in the $0$ degree configuration. The applied force is that of the weight of the thermal bus itself as well as of thermal straps hanging off the thermal bus. The red surfaces indicate the fixed contacts in the simulation. The maximum axial load was 170\,N, which occurred in the $45$\,degree simulation. }
\label{fig:sim}
\end{figure}

In support of this truss design, we performed mechanical simulations using SolidWorks.\footnote{Dassault Systèmes SolidWorks Corporation, 175 Wyman Street, Waltham, MA 02451, USA} The purpose of these simulations was to estimate the maximum axial load along the carbon fiber rods in a variety of orientations, as well as to ensure that the carbon fiber rods themselves would not buckle under load. The maximum axial load allows us to estimate the maximum pulling force on the strut glue joints, which in turn gives us the ability to estimate a factor of safety for those glue joints, in conjunction with the pull testing we performed. We used a model, shown in Figure~\ref{fig:sim}, consisting of the $100$\,mK thermal BUS and its support trusses, with slight simplifications of irrelevant geometry, such as removing bolt holes. The $1$\,K ends of the struts were fixed, and a bonded global contact was used. Each of the carbon fiber rods was treated as a strut. The thermal BUS material was set to the SolidWorks default copper, and the strut feet were set to the SolidWorks default aluminum 6061. The carbon fiber rod material properties were sourced from vDijk Pultruded Products.\footnote{Aphroditestraat 24, NL-5047, TW TILBURG, The Netherlands, http://www.dpp-pultrusion.com/en/the-company/} While not the same brand of carbon fiber that we used in our struts, the minimum factor of safety of the carbon fiber rods across all simulations was $>100$, so slight variations in the material properties are negligible, and critically have no bearing on the modeled axial force. The applied load was the weight of the system, $45$\,kg, and it was applied in $5$ orientations: $-90$, $-45$, $0$, $45$, and $90$ degrees with respect to the plane of the thermal bus. We performed both static and buckling simulations. Across all orientations, the minimum factor of safety on the failure of the carbon fiber itself (excluding the glue joint) was 117. The maximum axial load on a carbon fiber strut was 170 N; requiring a factor-of-safety (FoS) of 5 gives a maximum pull force of $850$\,N. Therefore, the glue joints must be able to withstand $850$\,N of axial load before failing. 


\section{Testing Method}
\label{sec:testing}
To test the performance of the glue joint, we pulled struts using a $50$\,kN load cell. The geometry of our struts, specifically the angle that the carbon fiber tube makes to the mating surface of the feet, makes them difficult to pull test directly. As such, we made test struts which exactly replicated the glue joint used in our struts, but which had the carbon fiber rod perpendicular to the mating interface, simplifying pull testing. For each strut we tested, we pulled it to failure at a rate of $0.02$ inches per minute.


\begin{figure}
\centering
\includegraphics[width=\columnwidth, trim={0.25cm 0.25cm 0cm 0.25cm}, clip]{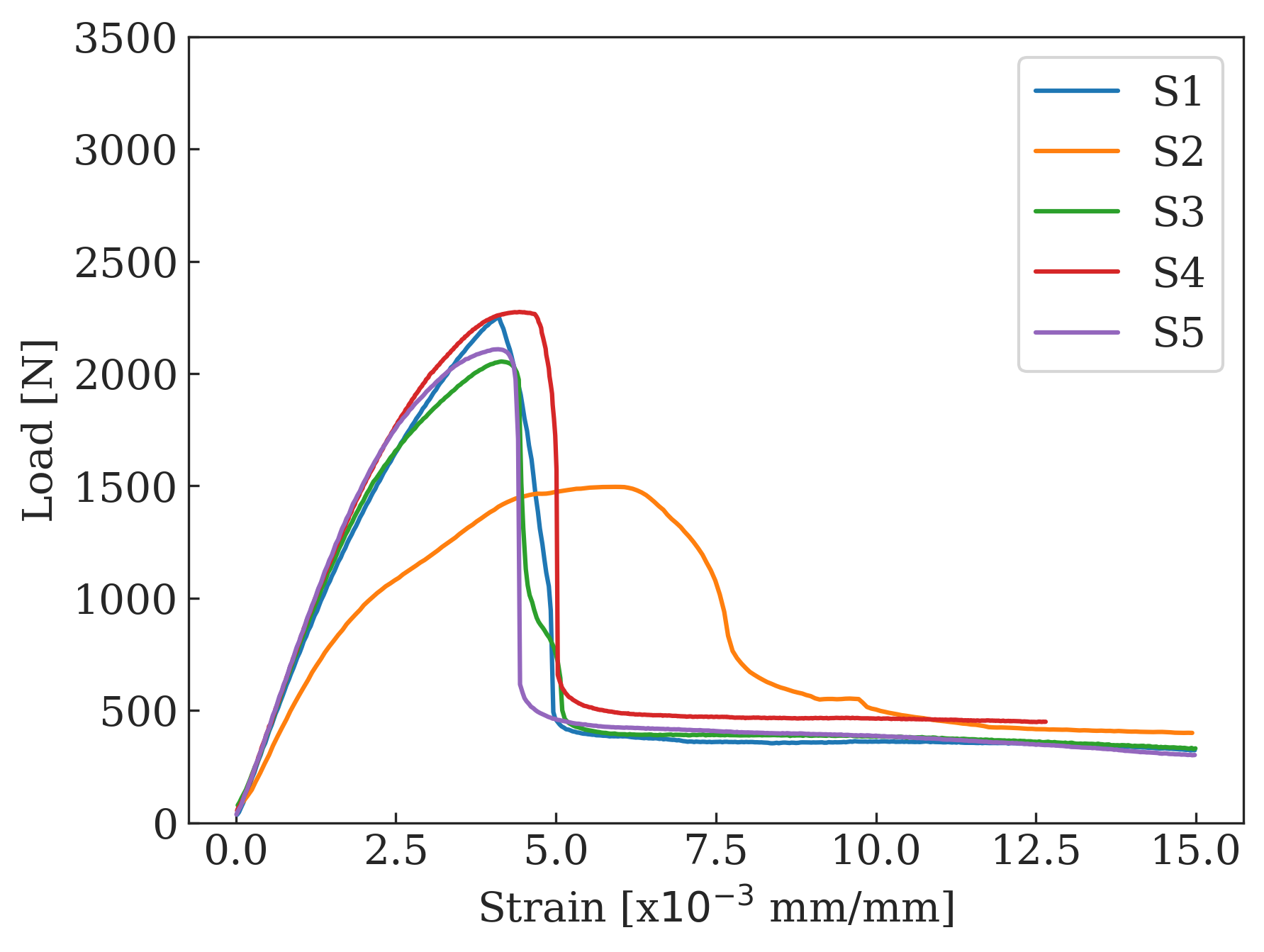}
\caption{Strain/load diagram for the smooth-walled strut design. Note the rapid ultimate failure of the struts at $\sim 4\times 10^{-3}$\,mm/mm after entering the plastic regime at $\sim 2.510^{-3}$\,mm/mm.}
\label{fig:smooth}
\end{figure}



\begin{figure}
\centering
\includegraphics[width=\columnwidth, trim={0.25cm 0.25cm 0cm 0.25cm}, clip]{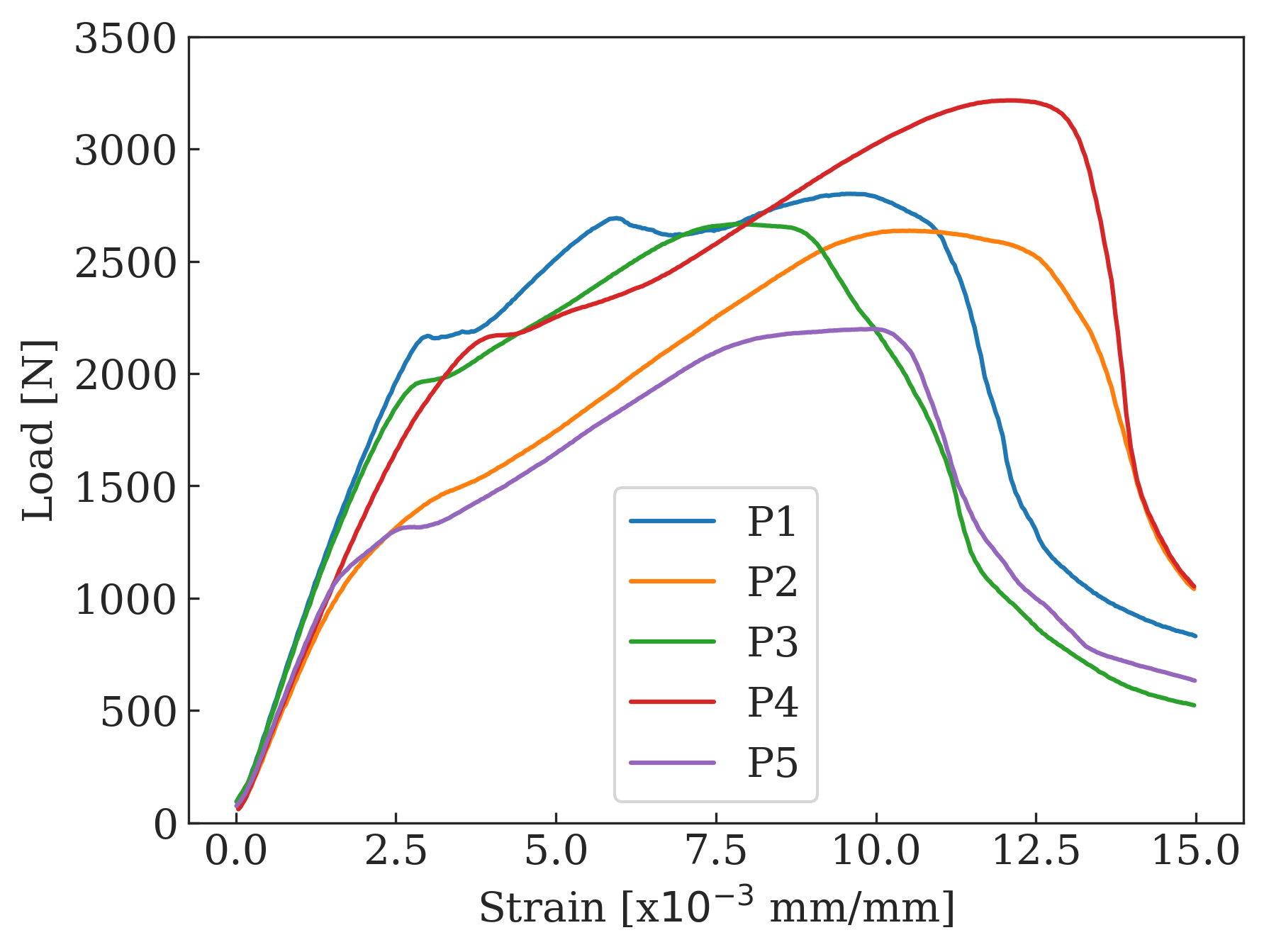}
\caption{Strain/load diagram for the profile-walled strut design. Note the long plateau prior to ultimate failure of the struts at $\sim 8\times 10^{-3}$\,mm/mm after entering the plastic regime at $\sim 2.510^{-3}$\,mm/mm, even for samples B2 and B5, which had a lower yield strength.}
\label{fig:profile}
\end{figure}


\section{Results}
\label{sec:results}
\begin{figure}
\centering
\includegraphics[width=\columnwidth]{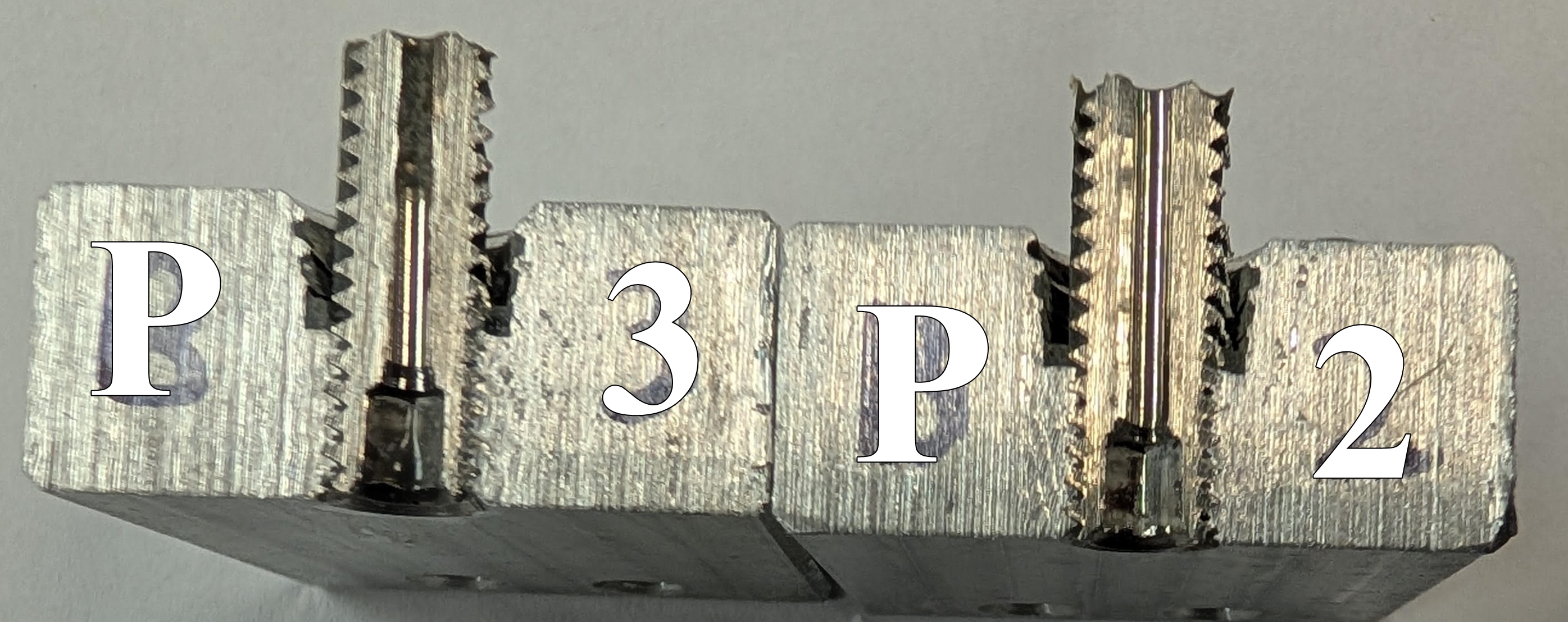}
\caption{Side-by-side of P3 (left) and P2 (right) cut in half through the failed glue joint. Visual inspection reveals no obvious difference in the glue joint, in contrast to S2, where there was a visiual difference between S2 and the other smooth-walled samples. }
\label{fig:p2p3}
\end{figure}

The results of this pull testing are shown in Figures~\ref{fig:smooth} and~\ref{fig:profile}. We tested five samples of each of the two interface designs, for a grand total of 10 struts. For the smooth-walled struts, the average ultimate strength was $2040\pm 280$N, while for the profile-walled struts, the average ultimate strength was $2700 \pm 330$N, where the uncertainties here are the standard deviations. The average yield strength for the smooth-walled samples was $1550 \pm 250$N while for the profile-walled, it was $1720\pm 430$N. Yield strengths were computed using the $0.2\%$ offset method, although due to the very low strain, the chosen offset was $0.02\%$. The full results are given in Table~\ref{tab:results}. Both the smooth and profile-walled struts met our performance requirement of $850$\,N, although the profile-walled struts performed better, particularly in ultimate strength. For all of the struts, the glue adhesively failed at the inner wall of the carbon fiber tube, as well as the bottom. Since the bottom and inner surface does not have a laminate, we did not roughen them with sandpaper. For all of the smooth-walled struts, the glue had adhesively failed at the outer aluminum wall: in other words, with roughening of the carbon fiber the adhesion of the glue to the carbon fiber surface was stronger that to the aluminum. For the profile-walled struts, the glue on the aluminum surfaces cohesively failed. Of the smooth-walled struts, there was one anomalously poorly performing strut, S2. On inspection, glue was still adhered to the bottom S2, unlike all the other smooth-walled struts where the glue had adhesively failed at the bottom of the tube. This suggests that there may have been an air bubble at the bottom of the groove, such that the glue never adhered to the aluminum surface, weakening it. P2 and P5 also showed significantly worse performance than the other profile-walled samples. Unfortunately, visual inspection did not reveal the source of this performance discrepancy. We cut samples P2 and P3 in half at the glue joint which failed; the results can be seen in Figure~\ref{fig:p2p3}. To the authors' eye there is no difference in the glue joint. Note that voide around the set screw is where the carbon fiber leg pulled out; there is no glue gap in either sample when the leg is inserted. 


\begin{table}[]
    \centering    
    \begin{tabular}{cccc}
    \hline\hline\noalign{\smallskip}
        Sample  & Yield strength   & Ultimate strength & Elongation at\\
                &         (N)            &         (N)             &Yield (mm/mm)\\
        \hline
        S1 & 1700 & 2300 & 0.0026 \\
        S2 & 1100 & 1500 & 0.0023\\
        S3 & 1600 & 2100 & 0.0024\\
        S4 & 1700 & 2300 & 0.0024\\
        S5 & 1700 & 2100 & 0.0024\\
        P1 & 2200 & 2800 & 0.0029\\
        P2 & 1200 & 2700 & 0.0022\\
        P3 & 2000 & 2700 & 0.0027\\
        P4 & 2100 & 3200 & 0.0035\\
        P5 & 1200 & 2200 & 0.0019\\
    \noalign{\smallskip}
    \hline
    \end{tabular}
    \label{tab:results}
    \caption{Yield and ultimate strengths, as well as elongation, for the $10$ struts in our study. Samples labeled ``S'' have a smooth-walled glue joined, while those labeled ``P'' have a profile-wall. The average ultimate strength for the smooth-walled struts is $2000 \pm 300$N, while that for the profile-walled is $2700 \pm 300$N, and the respective yield strengths are $1600 \pm 300$N and $1700\pm 400$N. }
\end{table}



\section{Conclusions and Improvements}
\label{sec:conc}
The profile-walled struts on average had both a greater yield and ultimate strength than the smooth-walled struts. The ultimate strength at failure for every profile-walled strut is higher than the ultimate strength at failure for every smooth-walled strut. The yield strengths for the profile-walled struts did not show consistent improvement with respect to those of the smooth-walled struts, although on average the performance was improved. There is also significantly more variation in the strain/load diagrams for the profile-walled struts than for the smooth-walled. It is possible that variation in the amount of glue worked into the thread contributed to this variation, although inspection of the failed glue joints of P2 and P3 did not reveal any obvious difference. Finally, the failure performance of the struts changed. While the smooth-walled struts generally failed shortly after reaching their yield strength, the profile-walled struts have significantly longer plateaus to their ultimate strength. While struts reaching yield strength need to be replaced, the high ultimate yield of the profile-walled struts gives a grace period before catastrophic failure. \\
It should be emphasized that the ultimate and yield strengths reported here are specific to the construction geometry that we have chosen and should not be extrapolated from. The salient results are that the profile-walled glue interface design improves the ultimate and yield strength of the joint as compared to the smooth-walled design, and that factor of safety of the profile-walled struts is $7$ assuming the worst yield strength of $1200$N and $10$ assuming the average yield strength. \\
In addition to the in-lab testing, these struts have been deployed in the SO LATR since early 2022, and have undergone numerous cool-downs from $300$\,K to $100$\,mK while under load. Each time the LATR has been opened up, most recently January of 2026, they have been inspected for damage. So far no damage has been observed to either the carbon fiber or to the glue joint.\\
While the failure mode for our design is the cohesive failure of the glue, the force at which the glue cohesively fails can in principle still be improved by varying the geometry of the wall profile. This is easily achieved by varying the size, pitch, etc. of the tap and set screw. We did not pursue this avenue of improvement as our struts were sufficiently strong with the profiles selected initially. Additionally, it was very difficult to roughen the walls of the feet and the set screws with sandpaper owing to their geometry. Sandblasting those surfaces would increase the adhesive strength of the glue to those surfaces, potentially increasing the strength of the join if the adhesive failure mode were to become dominant. On the other hand, it was evident that very little glue contact was made between the glue and the inner wall of the carbon fiber tube with either strut design. Roughening the inside of the carbon fiber tube in addition to the outside may improve this, as well as pre-applying glue to the inner surface of the carbon fiber tube before inserting it into the joint. 

\section{Acknowledgments}
They authors would like the acknowledge the Penn Engineering department for graciously proving access to their load cell for this work and to Peter Bruno in generously teaching us to operate it. This work was supported in part by a grant from the Simons Foundation (Award Number: 457687, B.K.). This work was supported in part by a grant from the Simons Foundation (Award Number: 457687, B.K.). This work was supported by the U.S. National Science Foundation (Award Number:  2153201).


\bibliography{references}


\end{document}